\documentclass[11pt,twoside]{article}
\usepackage{./asp2014}
\usepackage{comment}

\usepackage{float}

\usepackage{array}
\newcolumntype{L}[1]{>{\raggedright\let\newline\\\arraybackslash\hspace{0pt}}m{#1}}
\newcolumntype{C}[1]{>{\centering\let\newline\\\arraybackslash\hspace{0pt}}m{#1}}
\newcolumntype{R}[1]{>{\raggedleft\let\newline\\\arraybackslash\hspace{0pt}}m{#1}}

\aspSuppressVolSlug
\resetcounters

\begin{document}

\title{The ngVLA Reference Design}
\author{Robert J. Selina,$^1$ Eric J. Murphy,$^2$ Mark McKinnon,$^1$ Anthony Beasley,$^2$ Bryan Butler,$^1$ Chris Carilli,$^1$ Barry Clark,$^1$ Steven Durand,$^1$ Alan Erickson,$^1$ Rafael Hiriart,$^1$ Wes Grammer,$^1$ James Jackson,$^1$ Brian Kent,$^2$ Brian Mason,$^2$ Matthew Morgan,$^2$ Omar Yeste Ojeda,$^2$ Viviana Rosero,$^1$ William Shillue,$^2$ Silver Sturgis,$^1$ and Denis Urbain$^1$}
\affil{$^1$National Radio Astronomy Observatory, Socorro, NM, USA}
\affil{$^2$National Radio Astronomy Observatory, Charlottesville, VA, Country}

\paperauthor{Robert J. Selina}{rselina@nrao.edu}{ORCID_Or_Blank}{National Radio Astronomy Observatory}{ngVLA}{Socorro}{NM}{87801}{USA}
\paperauthor{Eric J. Murphy}{emurphy@nrao.edu}{ORCID_Or_Blank}{National Radio Astronomy Observatory}{ngVLA}{Charlottesville}{VA}{22903}{USA}
\paperauthor{Mark McKinnon}{mmckinno@nrao.edu}{ORCID_Or_Blank}{National Radio Astronomy Observatory}{ngVLA}{Socorro}{NM}{87801}{USA}
\paperauthor{Anthony Beasley}{tbeasley@nrao.edu}{ORCID_Or_Blank}{National Radio Astronomy Observatory}{ngVLA}{Charlottesville}{VA}{22903}{USA}
\paperauthor{Bryan Butler}{bbutler@nrao.edu}{ORCID_Or_Blank}{National Radio Astronomy Observatory}{ngVLA}{Socorro}{NM}{87801}{USA}
\paperauthor{Chris Carilli}{ccarilli@nrao.edu}{ORCID_Or_Blank}{National Radio Astronomy Observatory}{ngVLA}{Socorro}{NM}{87801}{USA}
\paperauthor{Barry Clark}{bclark@nrao.edu}{ORCID_Or_Blank}{National Radio Astronomy Observatory}{ngVLA}{Socorro}{NM}{87801}{USA}
\paperauthor{Steven Durand}{sdurand@nrao.edu}{ORCID_Or_Blank}{National Radio Astronomy Observatory}{ngVLA}{Socorro}{NM}{87801}{USA}
\paperauthor{Alan Erickson}{aerickso@nrao.edu}{ORCID_Or_Blank}{National Radio Astronomy Observatory}{ngVLA}{Socorro}{NM}{87801}{USA}
\paperauthor{Wes Grammer}{wgrammer@nrao.edu}{ORCID_Or_Blank}{National Radio Astronomy Observatory}{ngVLA}{Socorro}{NM}{87801}{USA}
\paperauthor{Rafael Hiriart}{rhiriart@nrao.edu}{ORCID_Or_Blank}{National Radio Astronomy Observatory}{ngVLA}{Socorro}{NM}{87801}{USA}
\paperauthor{James Jackson}{jjackson@nrao.edu}{ORCID_Or_Blank}{National Radio Astronomy Observatory}{ngVLA}{Socorro}{NM}{87801}{USA}
\paperauthor{Brian Kent}{bkent@nrao.edu}{ORCID_Or_Blank}{National Radio Astronomy Observatory}{ngVLA}{Charlottesville}{VA}{22903}{USA}
\paperauthor{Brian Mason}{bmason@nrao.edu}{ORCID_Or_Blank}{National Radio Astronomy Observatory}{ngVLA}{Charlottesville}{VA}{22903}{USA}
\paperauthor{Matthew Morgan}{mmorgan@nrao.edu}{ORCID_Or_Blank}{National Radio Astronomy Observatory}{ngVLA}{Charlottesville}{VA}{22903}{USA}
\paperauthor{Omar Ojeda}{oojeda@nrao.edu}{ORCID_Or_Blank}{National Radio Astronomy Observatory}{ngVLA}{Charlottesville}{VA}{22903}{USA}
\paperauthor{Viviana Rosero}{vrosero@nrao.edu}{ORCID_Or_Blank}{National Radio Astronomy Observatory}{ngVLA}{Socorro}{NM}{87801}{USA}
\paperauthor{William Shillue}{bshillue@nrao.edu}{ORCID_Or_Blank}{National Radio Astronomy Observatory}{ngVLA}{Charlottesville}{VA}{22903}{USA}
\paperauthor{Silver Strugis}{ssturgis@nrao.edu}{ORCID_Or_Blank}{National Radio Astronomy Observatory}{ngVLA}{Socorro}{NM}{87801}{USA}
\paperauthor{Denis Urbain}{durbain@nrao.edu}{ORCID_Or_Blank}{National Radio Astronomy Observatory}{ngVLA}{Socorro}{NM}{87801}{USA}

\begin{abstract}
The next-generation Very Large Array (ngVLA) is an astronomical observatory planned to operate at centimeter wavelengths (25 to 0.26 centimeters, corresponding to a frequency range extending from 1.2 to 116\,GHz). The observatory will be a synthesis radio telescope constituted of approximately 244 reflector antennas each of 18 meters diameter, and 19 reflector antennas each of 6 meters diameter, operating in a phased or interferometric mode.
We provide a technical overview of the Reference Design of the ngVLA. This Reference Design forms a baseline for a technical readiness assessment and the construction and operations cost estimate of the ngVLA. The concepts for major system elements such as the antenna, receiving electronics, and central signal processing are presented.
\end{abstract}

\section{Introduction}

As part of its mandate as a national observatory, the National Science Foundation's (NSF) National Radio Astronomy Observatory (NRAO) is looking toward the long-range future of radio astronomy and fostering the long-term growth of the U.S. and global astronomical community. With NSF support, NRAO has sponsored a series of science and technical community meetings to define the science mission and concept for a next-generation Very Large Array \citep[ngVLA;][]{1} that builds on the legacies of the Atacama Large Millimeter/submillimeter Array (ALMA) and the Jansky Very Large Array (VLA).

Based on input solicited from the astronomical community, the ngVLA is planned as an astronomical observatory that will operate at centimeter wavelengths (25 to 0.26 centimeters, corresponding to a frequency range extending from 1.2 to 116\,GHz). The observatory will be a synthesis radio telescope consisting of:
\begin{itemize}
    \vspace{-6pt}
    \item A main array of 214 reflector antennas each of 18 meters diameter, operating in a phased or interferometric mode. The main array is distributed to sample a wide range of scales from 10s of meters to 1000 km. A dense core and spiral arms provide high surface brightness sensitivity, with mid-baseline stations enhancing angular resolution. 
    \vspace{-6pt}
    \item A short baseline array (SBA) of 19 reflector antennas of 6\,m aperture will be sensitive to a portion of the larger angular scales undetected by the main array.  The SBA may be combined with 4 18\,m (main-array) antennas used in a total power mode to completely fill in the central hole in the ($u,v$)-plane left by the 6\,m dishes.
    \vspace{-6pt}
    \item A long baseline array (LBA) will add an additional 30 reflector antennas each of 18\,m diameter in 10 clusters providing continental scale baselines ($B_{MAX} \sim 8860\,$km). The LBA is designed to sample a broad range of scales for stand-alone sub-array use, as well as for integrated operation with the main array. 
\end{itemize}
It total, the ngVLA will have approximately ten times the sensitivity of the VLA and ALMA, continental-scale baselines providing sub-milliarcsecond-resolution, and a dense core on km-scales for high surface brightness sensitivity. Such an array bridges the gap between ALMA, a superb sub-mm array, and the future SKA1, optimized for longer wavelengths.

The dense core and the signal processing center of the array will be located at the Very Large Array site, on the plains of San Agustin, New Mexico. The high desert plains of the Southwest U.S., at over 2000\,m elevation, provide excellent observing conditions for the frequencies under consideration, including reasonable phase stability and opacity at 3\,mm wavelength over a substantial fraction of the year.

The array will also include stations in other locations throughout the state of New Mexico, west Texas, eastern Arizona, and northern Mexico. Long baseline stations are located in Hawaii, Washington, California, Iowa, Massachusetts, New Hampshire, Puerto Rico, the US. Virgin Islands, and Canada. 

Operations will be conducted from both the VLA Site and the Array Operations and Repair Centers in Socorro, NM. A Science Operations Center and Data Center are collocated in a large metropolitan area and will be the base for science operations and support staff, software operations and related administration. Research and development activities will be split amongst these centers as appropriate.

The facility will be operated as a proposal-driven instrument. The fundamental data products delivered to ngVLA users will be science-ready data products (i.e., images and cubes) generated using calibration and imaging pipelines created and maintained by the project. Both the pipeline products and the ``raw'' visibilities and calibration tables will be archived, retaining the option of future re-processing and archival science projects. 

The ngVLA project is developing a Reference Design for the array as a baseline for construction and operation costing, and future design trade-off decisions. This Reference Design is intended to be low technical risk in order to provide a degree of conservativism in the estimates. However, leading-edge concepts and techniques that may improve the performance and/or reduce cost are being developed in parallel, and will be evaluated in the conceptual design phase of the facility. This paper provides an overview of the Reference Design as the project approaches the Astro2020 Decadal Survey.

\section{Key Science \& Technical Requirements}

The Key Science Goals and all other science use cases were parameterized and analyzed \citep{12} to determine the science requirements for the ngVLA \citep{4}. While this aspect of the requirements definition is top-down and mission-driven, some judicious adjustment of the requirements is still appropriate. A primary science requirement for the ngVLA is to be flexible enough to support the breadth of scientific investigations that will be proposed by its creative scientist-users over the decades-long lifetime of the instrument. The requirements have therefore been adjusted to provide a balanced, flexible, and coherent complement of capabilities. The primary requirements that drive the design are described below:   

\begin{itemize}

\item \textbf{Frequency Coverage:} The ngVLA should be able to observe in all atmospheric windows between 1.2 and 116\,GHz. These frequency limits are bracketed by spectral line emission from H{\sc i} and CO respectively.

\item \textbf{Continuum Sensitivity:} A continuum sensitivity of better than 0.02 $\mu$Jy/bm at 30\,GHz and 0.2 $\mu$Jy/bm 100\,GHz is required for studying protoplanetary disks. This requires a combination of large collecting area and wide system bandwidth. 

\item \textbf{Line Sensitivity:} A line sensitivity of 30 $\mu$Jy/bm/km/s for frequencies between 10 and 50\,GHz is simultaneously required to support both astrochemistry studies and deep/blind spectral line surveys. A line sensitivity of $1 - 100$\,mK at $0.1 - 5\arcsec$ angular resolution and $1 - 5$\,km/s spectral resolution between 70 and 116\,GHz is required to simultaneously support detailed studies of CO and variations in gas density across the local universe. The spectral line cases push the system design towards quantum-limited noise performance at the expense of bandwidth above 10\,GHz. 

\item \textbf{Angular Resolution:} A synthesized beam having a FWHM better than 5\,mas with uniform weights is required at both 30 and 100\,GHz, while meeting the continuum sensitivity targets.

\item \textbf{Largest Recoverable Scale:} Angular scales of $>20\arcsec \times (100\,{\rm GHz}/\nu)$ must be recovered at frequencies $\nu < 100$\,GHz. A more stringent desire is accurate flux recovery on arcminute scales at all frequencies. These scales approach the size of the primary beam of an 18\,m dish, so both shorter baselines and a total power capability are necessary to completely fill in the central hole in the ($u,v$)-plane.  

\item \textbf{Surface Brightness Sensitivity:} The array must provide high-surface brightness sensitivity over the full range of angular scales recoverable with the instrument. This leads to a centrally condensed distribution of antennas.

\item \textbf{Brightness Dynamic Range:} The system brightness dynamic range shall be better than 50\,dB for deep field studies. This requirement pushes a number of systematic requirements including pointing, gain, and phase stability. 

\item \textbf{Survey Speed:} The array shall be able to map a $\sim$10 square degree region to a depth of $\sim 1\,\mu$Jy/bm at 2.5\,GHz and a depth of $\sim 10\,\mu$Jy/bm at 28\,GHz within a 10 hr epoch for localization of transient phenomena identified with other instruments. Holding collecting area and receiver noise constant, this favors smaller apertures. 

\item \textbf{Beamforming for Pulsar Search, Pulsar Timing and VLBI:} The array shall support no less than 10 beams spread over 1 to 10 subarrays that are transmitted, over the full available bandwidth, to a VLBI recorder/correlator, pulsar search engine or pulsar timing engine. The pulsar search and timing engine must be integral to the baseline design. 

\item \textbf{Science Ready Data Products:} The primary data product delivered to users shall be calibrated images and cubes. Uncalibrated, ``raw" visibilities shall be archived to permit reprocessing. Producing these higher-level data products requires some standardization of the initial modes/configurations that the system is used in (e.g., limited tuning options), and repeatability/predictability from the analog system to reduce the calibration overheads. 

\end{itemize}

\section{Site Selection \& Performance}

The VLA site on the plains of San Agustin was originally chosen as the location for the array because of its desirable properties: large, relatively flat, undeveloped (to minimize RFI) yet not too remote (for accessibility), at low latitude (for sky coverage), and at high elevation (to minimize atmospheric effects) \citep{thompson1980}.  These properties still hold true, and motivate examination of the VLA site as the center of the ngVLA.  Furthermore, with extensive existing infrastructure, the VLA site leverages an already-existing system of power, fiber, and buildings, which will reduce cost.  The three main environmental or atmospheric quantities that may affect data, and what is known about them at the VLA site, are discussed in the following sections. 

\subsection{RFI}

The VLA site is remote enough that Radio Frequency Interference (RFI) is not a debilitating problem, so it will be possible to observe at the lower frequencies of ngVLA \citep{21}. Furthermore, the ngVLA will benefit by advanced studies of RFI detection and excision that are currently ongoing \citep{burnett2018}.  The degree of RFI characterization of the site reduces the risk in site selection, and leveraging existing infrastructure could create significant cost savings for both the construction and operation of the array.  Given the large extent of ngVLA ($B_{MAX} \sim 8860$ km), it is clear that the antennas which are outside the plains will experience different RFI environments than that at the site.  However, there are locations which are relatively free of locally generated RFI (downward RFI from orbiting satellites is ubiqitous and nearly site-independent), and the U.S. southwest has many such locations \citep{li2004}.

\subsection{Atmospheric Phase Stability}

\begin{figure}
\centering
\includegraphics[width=0.8\textwidth]{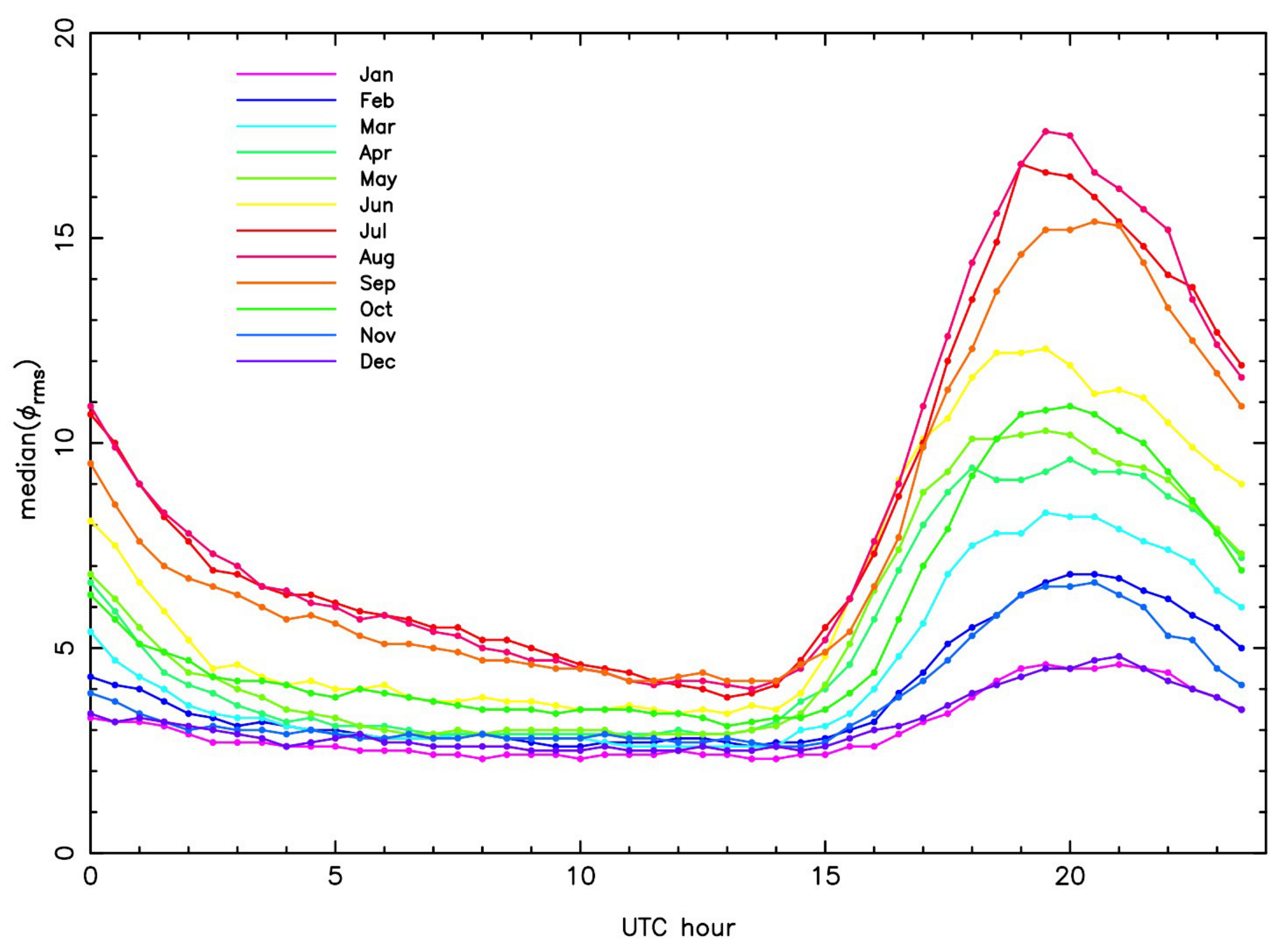}
\caption{\label{fig:rmsphase}The median rms phase measured with the atmospheric phase monitor at the VLA (300 m E-W baseline, 11.7 GHz beacon), from 1995 to 2017. Measurements are calculated over a 10 minute period after subtracting any linear trend.  Different months are plotted as different colors, as shown in the legend. }
\end{figure}

Analysis of data from the VLA site atmospheric phase monitor shows that fast switching phase calibration at 3\,mm should be viable for most of the year with a 30\,s total calibration cycle time \citep{19}.  This analysis was based on one year of atmospheric phase monitoring at the VLA site \citep{butler1999}.  A much longer time base of these values is now available.  Figure~\ref{fig:rmsphase} shows median values of the rms phase on the 300 m E-W baseline of the atmospheric phase monitor from 1995 through 2017, plotted as a function of UTC hour, and month.  It is easy to see that these fluctuations are small for much of the time, and only become greater than 10$^\circ$ (rms @ 11.7 GHz, over 10 minutes) in the summer during daytime.  Little information is available on phase fluctuations at  locations outside the plains; this is a topic to be studied to determine the ability to use the remote sites at the highest frequencies of ngVLA.  Note that there should also be a 25\,mJy calibrator source within 2$\deg$ in 98\% of observed fields, ensuring short slews.  Such a calibrator is adequate to ensure that the residual rms phase noise due to the signal-to-noise ratio on the phase calibrator is much less than that due to the troposphere, even for a 30\,s cycle time with only 3\,s on the calibrator each visit \citep{19,20}. The project is also investigating radiometric phase correction techniques as part of the ngVLA project to increase the total phase calibration cycle time.

\subsection{Atmospheric Opacity}

\begin{figure}[H]
\centering
\includegraphics[width=0.9\textwidth]{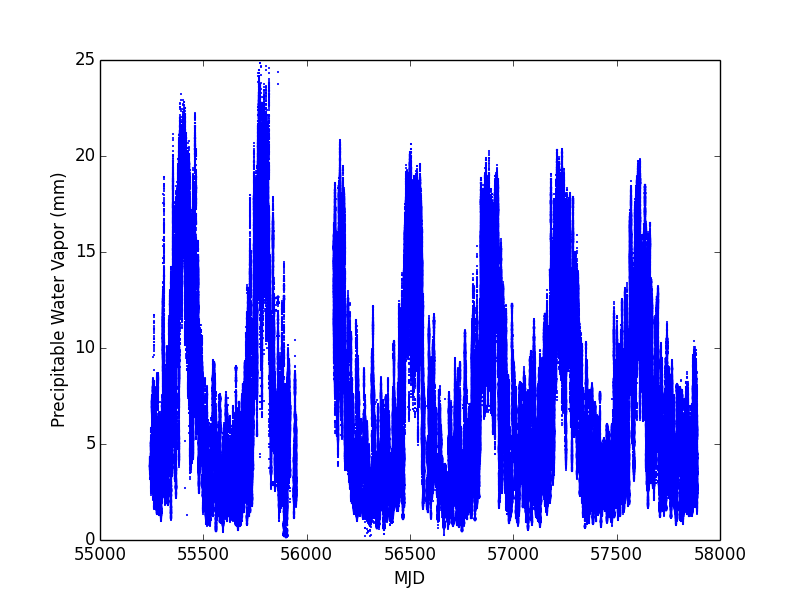}
\caption{\label{fig:pwv}PWV at the VLA site, estimated using surface weather measurements, from 2010 to 2017.  Note that a PWV of 6\,mm produces an opacity of less than 7\% at 90 GHz. }
\end{figure}

While at centimeter wavelengths atmospheric opacity is a relatively minor issue compared to phase stability, it becomes a much bigger issue at millimeter wavelengths.  Similar to the atmospheric phase stability data, there is a long-time baseline of surface weather data at the VLA site.  This can be used to estimate the atmospheric Precipitable Water Vapor (PWV), which is the main contributor to the fluctuating part of atmospheric opacity \citep{butler1998}.  Figure~\ref{fig:pwv} shows this value for the years 2010 through 2017.  In winter months, the median over all hours is around 3\,mm, and over the entire year the median over all hours is 5.4\,mm.  Vertical opacity for 5.4\,mm PWV at 90 GHz is less than 7\%, so opacity should not be a major problem for ngVLA.  As with RFI and phase stability, there is little information on atmospheric opacity at other locations, though it is almost always clear that higher sites have less opacity.  The project does have access to surface weather data, and to radiosonde launch data (twice per day) from NOAA for some tens of sites across the southwest US, which will be the subject of a future study to determine opacity properties across the extent of the ngVLA.

\subsection{Final Site Selection}

Because of the quality of the site for both low- and high-frequency observing, and the existing infrastructure, the ngVLA is centered near the current VLA.  The southwest U.S. and northern Mexico are sparsely populated and the antennas within 1000 km of the VLA are sited to select remote, radio quiet, and dry sites, while still considering the logistics of site access, electrical infrastructure and fiber optic network topology. The long baseline array sites were selected to minimize site impact and leverage shared infrastructure of other existing observatories, so sites operated by the VLBA or other observatories are preferred.  Note that the VLA site was used for acceptance testing of the original ALMA antennas, including observations up to 230\,GHz, and the experience was that the VLA site, at 2124\,m elevation is a high-quality 90\,GHz site - comparable to the Plateau de Bure site in overall performance \citep{22}.

\section{Array Configuration}

The ngVLA array design includes three fundamental subarrays  providing a wide range of angular scales: a main interferometric array, a short baseline array, and a long-baseline array.

\begin{figure}
\centering
\includegraphics[width=0.9\textwidth]{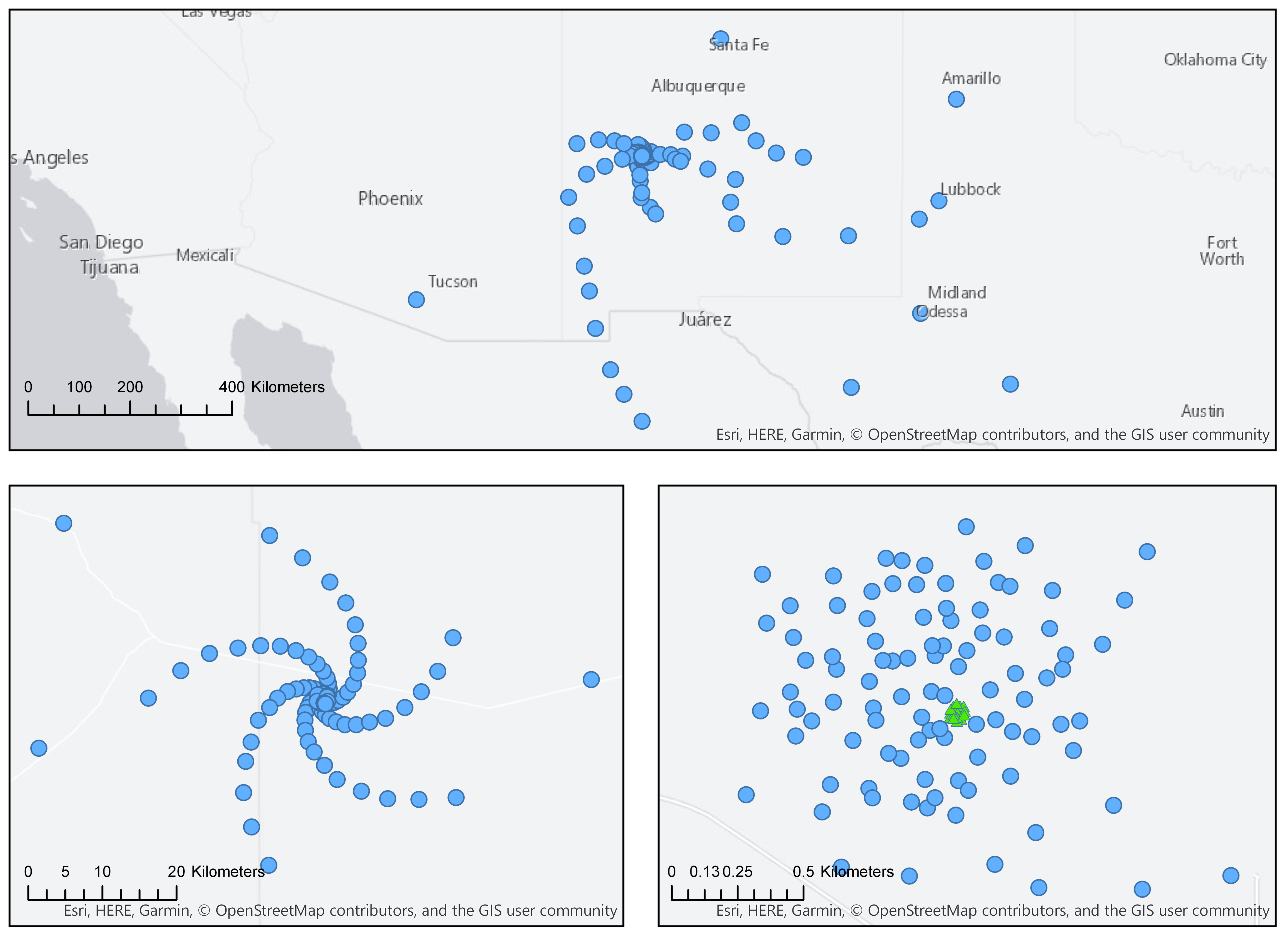}
\caption{\label{fig:config}Top: ngVLA Main Array Configuration Rev. B (Spiral-214). The antenna positions are still notional, but are representative for performance quantification and cost estimation. Bottom Left: Zoom view of the plains of San Agustin. Bottom Right: Zoom view of the compact core. SBA antennas are shown in green. }
\end{figure}

\begin{figure}
\centering
\includegraphics[width=0.9\textwidth]{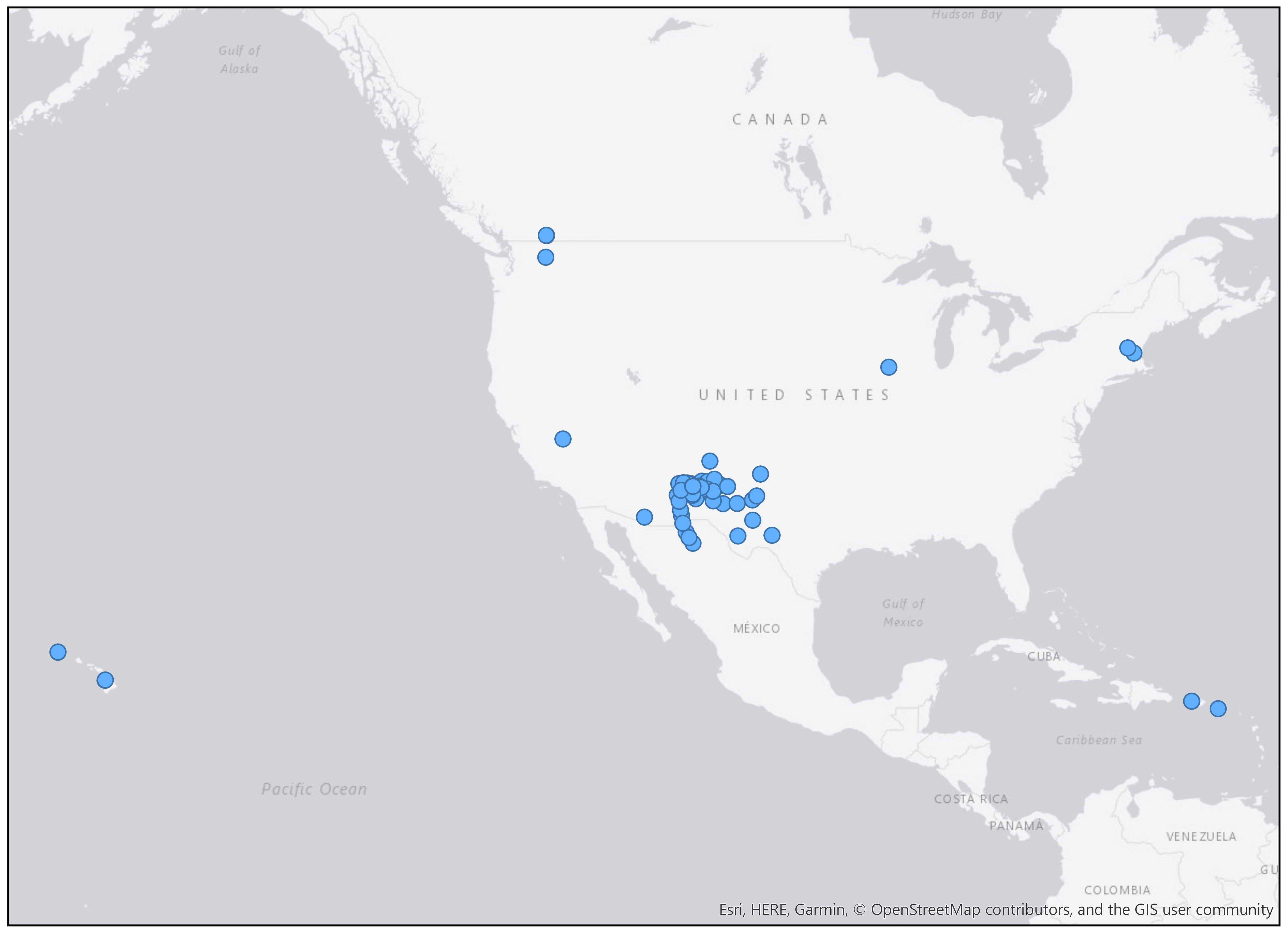}
\caption{\label{fig:LBAconfig}View of the Main Array and Long Baseline Array stations. Multiple antennas are located at each LBA site. }
\end{figure}

The main array configuration will consist of 214 18\,m antennas at the approximate locations shown in Figure  \ref{fig:config}. The array collecting area (see Table \ref{tab:arrayrad}) is distributed to provide high surface brightness sensitivity over a range of angular scales spanning from approximately 1000 to 100\,mas  while providing high point source sensitivity on scales up to 10\,mas. A large fraction of the collecting area is in a randomly distributed core to provide high snapshot imaging fidelity and there are arms extending asymmetrically out to \textasciitilde 1000 km baselines to fill the ($u,v$)-plane via Earth rotation and frequency synthesis. 

The design has been extended from the main interferometric array to include both a short spacing array and total power dishes \citep{16}. This was necessary after a review of the key science cases, as these are dependent on the recovery of large scale structure that approaches the size of the antenna primary beam. 

The auxiliary short baseline array (SBA) of 19 reflector antennas of 6\,m aperture will be sensitive to a portion of the larger angular scales undetected by the main array.  The SBA will provide spacings from $\sim$ 11\,m to 56\,m, providing comparable surface brightness sensitivity to the main array, in equal observing time, when the main array is ($u,v$)-tapered to the natural resolution of the SBA. This allows for commensal observing, and more importantly, full cross-correlation and cross-calibration of the SBA and main array. The array distribution is semi-randomized to improve the point spread function \citep{15}.  

The SBA will be combined with four 18\,m (main array) antennas used in total power (TP) mode to completely fill in the central hole in the ($u,v$)-plane left by the 6\,m dishes. It is a design goal to share the mount design of the 18\,m interferometric array antennas and the TP antennas, but this will require further study. 

In response to community feedback, a long baseline array (LBA) has also been added to the configuration (see Figure  \ref{fig:LBAconfig}). The long baseline array adds 30 antennas of 18\,m diameter at 10 additional sites. The LBA provides continental scale ($B_{MAX} \sim 8860$ km) baselines while also providing scales from $\sim$ 30 m to 1000 km within the subarray. This will enable the LBA to function effectively as a stand-alone array or as an integrated part of the main array. 

\begin{table}[t]
\centering
\begin{tabular}{c|c|c}
\hline
Radius	& Collecting Area Fraction & Quantity \\ 
\hline
\hline
~~~0\,km$ < R < $~~~1.3\,km	& $\approx 44$\% & 94 \\
1.3\,km$ < R < $~~~~36\,km	& $\approx 35$\% &  74\\
~36\,km$ < R < $1000\,km	& $\approx 21$\% & 46\\
\hline
\end{tabular}
\caption{\label{tab:arrayrad}Radial distribution of collecting area for the main array (214 antennas).}
\end{table}

\begin{table}[b]
\centering
\begin{tabular}{l|c|c|c|c}
\hline
Array Element	& Aperture Diameter	& Quantity	& $B_{MIN}$	& $B_{MAX}$	\\ 
                 	& [m]                      	&       	& [m]         	& [km]\\
\hline
\hline
Long Baseline Array 	& 18 	& 30 	& 32.6 & 8856 \\
Main Interferometric Array	& 18	& 214	& 30.6	& 1005 \\
Short Baseline Array	& 6	& 19	& 11.0 	& 0.06	\\
Total Power / Single Dish	& 18	& 4$^\dagger$	& -	& -	 \\
\hline
\end{tabular}
$^\dagger$These 4 dishes are included as part of the 214 main array.\hspace{3.25cm}
\caption{\label{tab:config}Summary of elements within the ngVLA array configuration.}
\end{table}

The ngVLA array configuration elements are summarized in Table \ref{tab:config}. The design of the array configuration is practical, accounting for logistical limitations such as topography, utility access,  local RFI sources and land management/availability.
An analysis of different weighting schemes (i.e., Briggs, ($u,v$)-taper) for specific science applications \citep{3} found that the current configuration provides a reasonable compromise and baseline for further iteration. The configuration will be a primary area of study in the coming years, e.g.,  investigations are underway to improve the imaging fidelity and quality of the synthesized beam.

\section{Array Calibration}

The calibration strategy for ngVLA is being developed early in the design so that it may guide the design of the hardware elements. The size and complexity of the calibration and imaging pipeline requires that the system design be responsive to its needs, and it should inform the design where possible. 

A secondary concern is the efficiency of the calibration process. Algorithms used must be suitable for parallel processing, antennas must not require much individual attention, and minimal human intervention should be generally required for routine operation. The calibration overheads applied will vary with the science requirements of a given observation, and less computationally or time intensive calibration approaches will be applied when possible. 

The operations plan calls for guaranteed time on source to each observer, with calibration overheads being the responsibility of the facility \citep{OpsCon}. This enables the reuse calibration observations for adjacent observations when their requirements are sufficiently similar, further improving observation efficiency. 

The general calibration strategies under consideration for the reference design are summarized below. 

\begin{itemize}
\item \textbf{Fast Atmospheric Phase Calibration:} Rapid atmospheric phase fluctuations will be mitigated by a combination of relative water vapor radiometry (WVR) and antenna switching cycles to astronomical phase calibrators. The switching cycle time will depend on empirical validation of the strategy, but is expected to be necessary on one to ten minute scales.  The antenna will be designed to both house the WVR and move 4$\deg$ on sky and settle to within the pointing specification with 10 seconds for elevation angles $<70\deg$ \citep{17}.  

\item \textbf{Slow Atmospheric \& Electronic Phase Calibration:} Slow atmospheric and electronic phase calibration will be achieved by traditional approaches, with astronomical phase calibrator observations bracketing all observations. Several astronomical calibrators may be used to map the slow varying terms, including ionospheric fluctuations. 

\item \textbf{Amplitude Calibration:} A list of known astronomical amplitude calibrators will be used to correct for system gain fluctuations within an observation and between observations taken over an extended period of time. The calibration pipeline will maintain a history of recent solutions to enable look-up of prior values. 

\item \textbf{Bandpass Calibration:} At a minimum, the system will correct for digital effects, given the predictable bandpass ripple from finite impulse response filters. The number of setups in the analog portions of the system will be limited, so typical calibration can also correct for analog bandpass effects based on historical look-up tables that are updated as the configuration of the system changes (i.e., when an antenna is serviced). 

\item \textbf{Polarization Calibration:} The use of linear feeds will require polarization calibration for most observations. Feeds may be placed at different (but known) position angles in the various antennas, so a single observation of a point source can solve simultaneously for the polarization leakage terms and the source polarization. Calibration for polarization as a function of position within the antenna beam will be assumed to be time invariant and corrected based on look-up tables for most observations.

\item \textbf{Relative Flux Density Calibration:} This calibration is used to tie together observations of a source taken over an extended period. The system will model atmospheric opacity based on barometric pressure and temperature monitored at the array core and each outlying station. A temperature stabilized noise diode will provide a flux reference, and when combined with corrections for modeled atmospheric opacity, we can assume a constant ratio in power from the switched noise calibrator and the source.

\item \textbf{Absolute Flux Density Calibration:} Absolute flux scale calibration will employ similar methods to relative calibration, with two notable changes. First, atmospheric tipping scans will be used to empirically determine atmospheric opacity, with improved fidelity. Second, observations of astronomical flux density calibrators will be used, along with the switched power system, to determine the absolute flux density of the source.
\end{itemize}

The ngVLA will need to maintain multiple lists of calibrators by calibration intent. The flux density calibrator list can be relatively small and based on the one built and maintained by the VLA. An extensive grid of sources will be required for phase and amplitude calibration. The large range of baselines present on the ngVLA means that it cannot be assumed that the source is unresolved at all scales, and the calibrators themselves will have to be imaged before use in the calibration process. 

\section{Antenna}

The antenna concept strikes a balance between competing science requirements and the programmatic targets for life cycle cost.  Sensitivity goals will be met, in part, by the total effective collecting area of the array. The reference design includes 244 antennas of 18\,m aperture (main array and long baseline array) and 19 antennas of 6\,m aperture (short baseline array) using an offset Gregorian optical design. 

The inclusion of frequencies down to 1.2\,GHz when combined with the operational cost targets significantly constrain the optical configuration. The use of feeds with wide illumination angles decreases their size such that they can be mounted within shared cryostats. This choice constrains the secondary angle of illumination to a degree that only Gregorian optical designs are practical. However, with a science priority of high imaging dynamic range in the $10-50$\,GHz frequency range, an offset Gregorian is near optimal.  The unblocked aperture will minimize scattering, spillover and sidelobe pickup. Maintenance requirements favor antenna optical configurations where the feed support arm is on the ``low side'' of the reflector.

The optimization for operations and construction cost suggests that a smaller number of larger apertures is preferable to larger numbers of small apertures. Survey speed requirements push the opposite direction, and a compromise value of 18\,m diameter is adopted for the reference design.  The design aims for Ruze performance to 116\,GHz, with a surface accuracy of 160 $\mu$m rms ($\lambda/16$ @ 116\,GHz) for the primary and subreflector combined under precision environmental conditions. The antenna optics are optimized for performance above 5\,GHz with some degradation in performance accepted at the lowest frequencies due to diffraction, in exchange for more stiffness in the feed arm to improve pointing performance.

Since the ngVLA is envisioned as a general purpose, proposal-driven, pointed instrument (rather than a dedicated survey telescope), the optics will be shaped to optimize the illumination pattern of single pixel feeds, increasing antenna gain while minimizing spillover. High pointing accuracy will also be necessary to provide the required system imaging dynamic range.  With an unblocked aperture, variations in the antenna gain pattern are expected to be dominated by pointing errors.  Preliminary requirements are for absolute pointing accuracy of 18 arc-seconds rms, with referenced pointing of 3 arc-seconds rms, during the most favorable environmental conditions \citep{16}.   

\begin{figure}[t]
\centering
\includegraphics[width=1.0\textwidth]{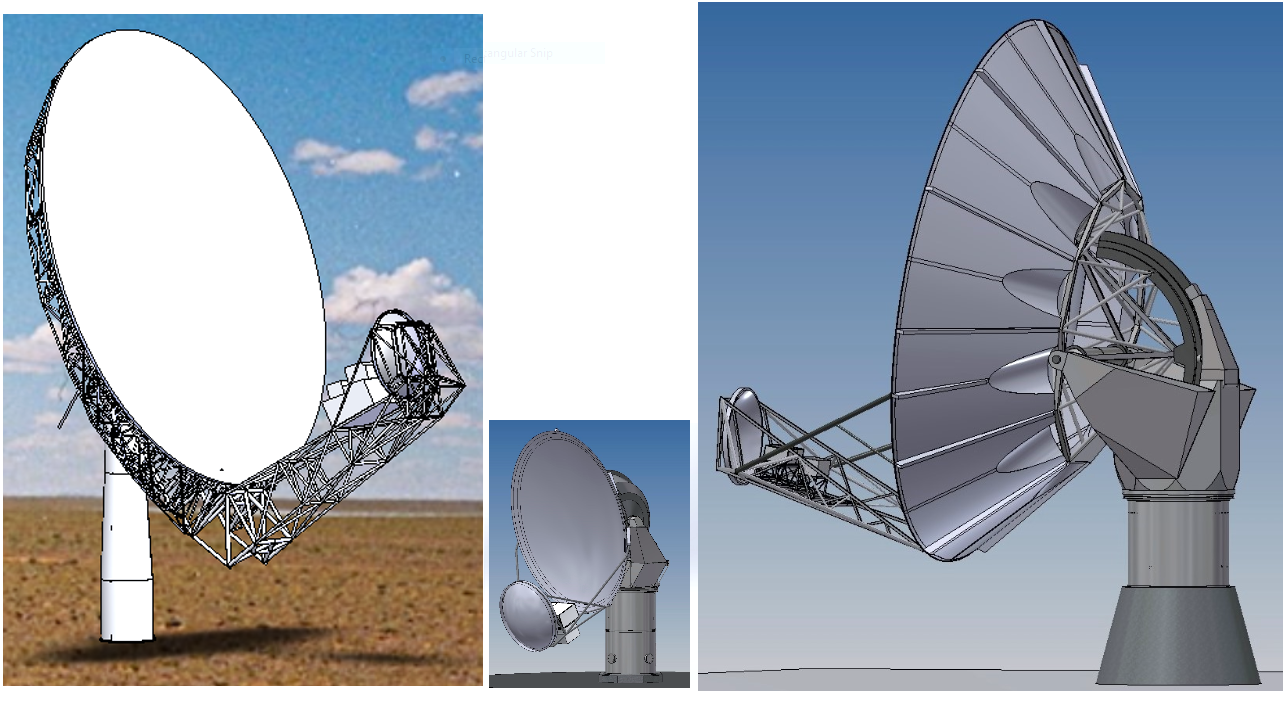}
\caption{\label{fig:antennas} Left: ngVLA 18\,m antenna reference design concept prepared by GDMS. Center: 6\,m short spacing array antenna concept prepared by NRCC.  Right: ngVLA 18\,m antenna composite design concept prepared by NRCC.}
\end{figure}

The mechanical and servo design is a typical altitude-azimuth design, Figure \ref{fig:antennas}.  Initial studies suggest pedestal designs are expected to have lower life-cycle cost while meeting pointing specifications. The antenna mechanical and servo design will be optimized for rapid acceleration and a fast settling time, in order to manage the switching overhead associated with short slews.

The project is pursuing a reference design to specifications for the 18\,m antenna with General Dynamics Mission Systems (GDMS). A parallel study into a composite design concept with the National Research Council of Canada (NRCC) is also underway, and NRCC are also preparing a reference design for the 6\,m short baseline array antenna.

The 6\,m aperture for the SBA antenna was chosen to provide overlapping ($u,v$)-plane coverage with the 18\,m dishes when the later are used in both interferometric and total power mode. The optical design is inherently more offset than the 18m design in order to maintain a suitable minimum subreflector aperture (2.7\,m) for 1.2\,GHz operation, but shares the same feed illumination angle. Maintaining common interfaces ensures that the 6\,m design can share the majority of the antenna electronics, including feeds and receivers. 

The 6\,m design employs a composite singe-piece reflector, and composite segmented backup structure on a steel pedestal mount. The mount includes space to house the digital electronics, power supplies and servo system, Figure \ref{fig:antennas}.

\section{Receiver Configuration}

The ngVLA will provide continuous frequency coverage from 1.2 - 50.5\,GHz and 70 - 116\,GHz in multiple bands.  Receivers will be cryogenically-cooled, with the receiver cryostats designed to integrate multiple receiver bands to the extent possible.  Limiting the number of cryostats will reduce both maintenance and electrical power costs.  The total number of bands required strongly depends on their fractional bandwidths: maximizing bandwidths will reduce the number of cryostats, with a possible penalty in sensitivity. Feeds for all receiver bands are cooled, and fully contained within the cryostat(s).

\begin{table}
\centering
\begin{tabular}{C{1.0cm}|C{1.0cm}|C{1.0cm}|C{1.0cm}|C{1.0cm}|C{1.0cm}|C{1.0cm}|C{1.0cm}}
\hline
Band & $f_L$ & $f_M$ & $f_H$ & BW & \multicolumn{3}{c}{Aptr. Eff., $\eta$A} \\
\# & (GHz) &  (GHz) & (GHz) & (GHz) & @$f_L$	& @$f_M$	& @$f_H$ \\ 
\hline
\hline
1 &1.2 &2.0 &3.5 &2.3 &0.80 &0.79 &0.74 \\
2	&3.5 &6.6 &12.3 &8.8 &0.80 &0.78 &0.76 \\
3	&12.3	&15.9	&20.5	&8.2	&0.84	&0.87	&0.86\\
4	&20.5	&26.4	&34	&13.5	&0.83	&0.86	&0.83\\
5	&30.5	&39.2	&50.5	&20	&0.81	&0.82	&0.78\\
6	&70	&90.1	&116	&46	&0.68	&0.61	&0.48\\
\hline
\end{tabular}
\caption{\label{tab:fe1}Band definitions and aperture efficiency of the baseline receiver concept. }
\end{table}

The baseline ngVLA receiver configuration consists of the low-frequency receiver (1.2 - 3.5\,GHz) in one cryostat, and five receivers spanning from 3.5 to 116\,GHz in a second cryostat. 
Bands 1 and 2 employ wideband feed horns and LNAs, each covering L$+$S bands, and C$+$X bands. Quad-ridged feed horns (QRFHs) are used, having dual coaxial outputs. Due to improved optical performance (improving illumination efficiency and reducing $T_{spill}$), cooled feeds, and the simplified RF design sensing linear polarization, the $T_{sys}$ is lower than current VLA L, S bands and comparable for C and X bands. Overall aperture efficiency and $T_{sys}$ are slightly degraded from optimal due to the wider bandwidths spanned, but this permits a compact package that can be affordably constructed and operated. 

The four high-frequency bands (12.3 - 116\,GHz) employ waveguide-bandwidth (\textasciitilde 1.67:1) feeds \& LNAs, for optimum aperture efficiency and noise performance. Axially corrugated feed horns with circular waveguide output ensure uniform illumination over frequency, with minimum spillover and resistive loss. 

\begin{table}
\centering
\begin{tabular}{C{1.0cm}|C{0.8cm}|C{0.8cm}|C{0.8cm}|C{0.8cm}|C{0.8cm}|C{0.8cm}|C{0.8cm}|C{0.8cm}|C{0.8cm}}
\hline
Band & \multicolumn{3}{c}{$T_{spill}$ (K)}	& \multicolumn{3}{c}{$T_{RX}$ (K)}	& \multicolumn{3}{c}{$T_{sys}$ (K)} \\
\# & @$f_L$	& @$f_M$	& @$f_H$	& @$f_L$	& @$f_M$	& @$f_H$	& @$f_L$	& @$f_M$	& @$f_H$ \\ 
\hline
\hline
1 	&12.8 	&10.1 	&4.0 	&9.9 	&10.3 	&13.8 	&27.1 &24.9 &22.4 \\
2	&12.8 	&7.0 	&3.9 	&13.4 	&15.4 	&14.4 	&30.8 &27.1 &23.6 \\
3	&4.1	&4.1	&4.1	&13.9	&16.9	&18.6	&23.3	&27.3	&36.3 \\
4	&4.1	&4.1	&4.1	&15.4	&16.2	&18.6	&33.1	&32.4	&36.0 \\
5	&4.1	&4.1	&4.1	&19.1	&20.4	&26.5	&34.0	&41.0	&101 \\
6	&4.1	&4.1	&4.1	&50.6	&49.0	&72.6	&123	&68	&189 \\
\hline
\end{tabular}
\caption{\label{tab:fe2}Noise performance of the baseline receiver concept. Assumes 1\,mm PWV for band 6, and 6\,mm PWV for others; 45$\deg$ elev. on sky for all. }
\end{table}

\begin{figure}[H]
\centering
\includegraphics[width=1.0\textwidth]{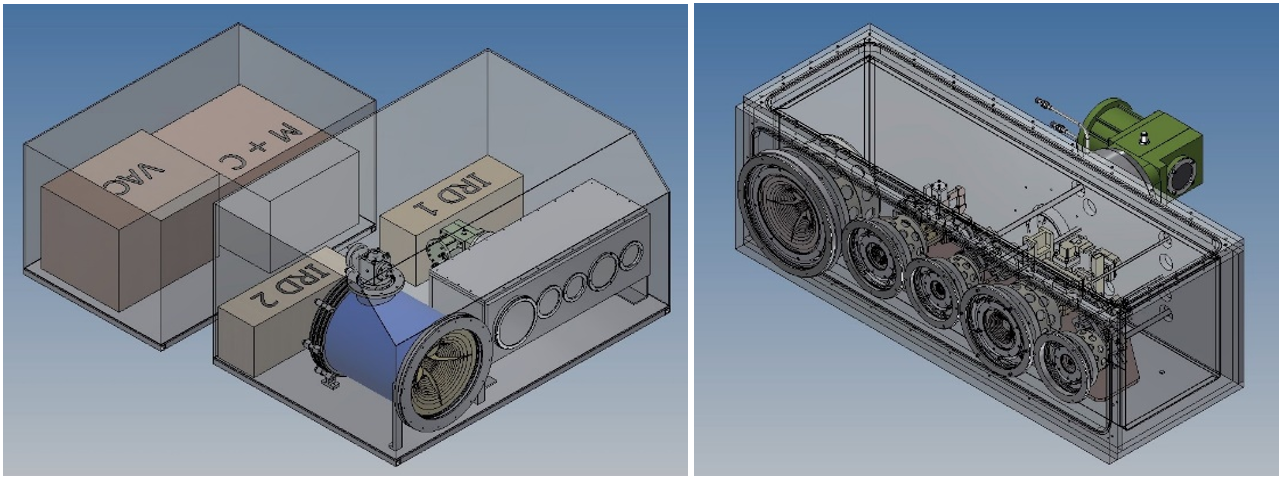}
\caption{\label{fig:frontend} Front end component packaging at the secondary focus of the antenna. Band selection and focus are achieved with a dual-axis translation stage.The integrated receiver packages (labeled IRD 1 and IRD 2) are located in close proximity to the cryostats. Bands $2-6$ are housed within in single cryostat.}
\end{figure}

\begin{figure}[H]
\centering
\includegraphics[width=1.0\textwidth]{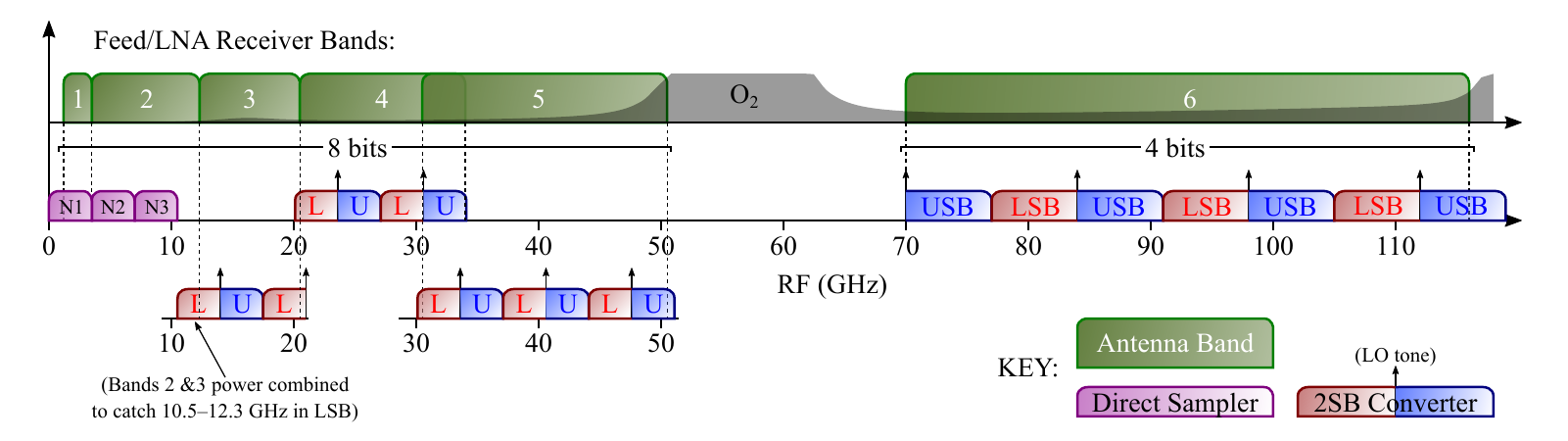}
\caption{\label{fig:sampler} Sampling concept employing integrated receiver technology for both direct and dual sideband converter/samplers. Direct single side-band 8-bit sampling is used for the first three Nyquist zones. Dual sideband 8-bit samplers are then used up to 50 GHz. The 70-116GHz band is spanned by 4-bit samplers, due to the reduced risk of persistent RFI at these frequencies.}
\end{figure}

The electronics concept relies on integrated receiver packages \citep{13} to further amplify  the  signals  provided  by  the  cryogenic  stage, down convert  them  if  necessary, digitize them, and deliver the resultant data streams by optical fiber to a moderately remote collection point (typically the antenna pedestal) where they can be launched onto a conventional network for transmission back to the array central processing facility.  Interfaces are provided for synchronization of local oscillators (LO's)  and  sampler  clocks,  power  leveling,  command  and  control,  health  and  performance monitoring, and diagnostics for troubleshooting in the event of component failure.

The integrated receiver concept is central to the antenna electronics concept for the ngVLA. Compact, fully-integrated,  field-replaceable, warm electronic modules support single-stage, direct-to-baseband downconversion (when needed), followed by a very low-power, low-overhead digitization scheme and an industry-standard fiber optic interface carrying unformatted serial data. The frequency plan is shown in Figure \ref{fig:sampler}.  

\section{Reference Distribution \& Data Transmission}

Given the large extent of the array, multiple time and frequency reference distribution concepts will be required to optimize for cost and performance. The array will be built as a combination of two different design methodologies.  

A large number of antennas are located on the Plains of San Agustin, and each of these will be connected directly to the central processing facility by dedicated, buried fiber optics. Roughly 70\% of the ngVLA antennas will be within this region. Clocks and local oscillator signals will be generated locally at the antenna and locked to a central reference with round trip phase correction. 

The remainder of the antennas, the mid and long baseline antennas, will fall into a VLBI\footnote{Very Long Baseline Interferometry.} station model with a number of local oscillator (LO) and data transmission stations located beyond the central core. These stations will be linked to the central timing system, correlator, and monitor and control system via long haul fiber optics. The most remote of these sites will have independent precision timing and frequency references, such as GPS-disciplined active hydrogen masers. Intermediate sites will used a mixed model dependent on the site logistics. 

\section{Central Signal Processor}

The Central Signal Processor (CSP) ingests the voltage streams recorded and packetized by the antennas and transmitted via the data transmission system, and produces a number of low-level data products to be ingested by the archive. Among its many functionalities, the CSP is responsible for compensating for the large transmission delays from the remote stations, correcting the I/Q channel imbalance of the receiver and improve the separation of the upper and lower sidebands, tracking the delay and phase differences between antennas, flagging the spectral channels corrupted with RFI at a pre-correlation stage, selecting the spectral window of interest within the digitized bandwidth, offsetting the different frequency standards used by the remote stations, and achieving the desired spectral resolution.

In addition to synthesis imaging, the CSP will support other capabilities required of modern telescopes to enable VLBI and time-domain science. The functional capabilities of the CSP include full-polarization auto- and cross-correlation computation, as well as beamforming capabilities for pulsar timing, pulsar/transient search, and VLBI recording. The CSP data products will vary by operation mode. The most common will be raw/uncalibrated visibilities, recorded in a common data model. The CSP will include all necessary ``back end'' infrastructure to average visibilities and package them for the archive, where they will be recorded to disk in a standard format. Calibration of these data products will be carried out through asynchronous data post-processing pipelines.

The CSP will support multiple sub-arrays operating simultaneously and fully independent from each other. Two key requirements for the system are the degree of commensality supported within a sub-array and the desired capabilities for sub-arrays operating simultaneously. At a minimum, the CSP will be able to compute auto- and cross-correlation products within a sub-array, as well as simultaneous cross-correlation and either pulsar timing, pulsar search or VLBI capabilities for different sub-arrays. Enabling correlation and beamforming products simultaneously within a sub-array is also under evaluation. Such a mode would reduce calibration overheads of the beamformer, and provide for localization/imaging concurrent with time-domain observations. The degree of commensality is expected to be a cost/complexity driver in the system and will be optimized on a best value basis.

The ngVLA correlator will employ an FX architecture, and will process an instantaneous bandwidth of up to 20\,GHz per polarization. The correlator-beamformer Frequency Slice Architecture \citep{14} developed by NRC Canada for the SKA Phase 1 mid-frequency telescope in South Africa \citep{SKA1_baseline} is well suited to ngVLA demands and is adopted for the reference design. This architecture will scale to the additional ngVLA apertures, bandwidth, and commensal mode requirements. Adopting this architecture could significantly reduce the non-recurring engineering costs during the design phase, while additional improvements in electrical efficiency can be expected from one additional FPGA manufacturing process improvement cycle due to ngVLA's later construction start date as compared to SKA Phase 1. Key performance requirements for the correlator are summarized in Table \ref{tab:csp}.

\begin{table}
\centering
\begin{tabular}{|L{5.8cm}|L{6.5cm}|@{}m{0pt}@{}} \hline
\textbf{Requirement Description} & \textbf{Specification} &\\[5pt] 
\hline
\hline
Number of Connected Antennas & 263 total &\\[5pt] \hline
Maximum Baseline Length	& 10,000\,km &\\[5pt] \hline
Maximum Instantaneous Bandwidth	& 20\,GHz per polarization &\\[5pt] \hline
Maximum Number of Channels	& $\geq 750,000$ channels  &\\[5pt] \hline
Highest Frequency Resolution	& 400\,Hz, corresponding to 0.1\,km/s resolution at 1.2\,GHz. &\\[17pt] \hline
Pulsar Search Beamforming	& \parbox[t]{6.5cm}{$\geq 10$ beams, \\ $\geq 1$\arcsec~coverage, 60\,km diameter sub-array} &\\[17pt]  \hline
Pulsar Timing Beamforming	& \parbox[t]{6.5cm}{$\geq 5$ independent sub-arrays\\ $\geq 1$ beam per sub-array} &\\[17pt]  \hline
\end{tabular}
\caption{\label{tab:csp}Central signal processor key specifications.}
\end{table}

\section {Post Processing System}

The software architecture for ngVLA will leverage NRAO's existing algorithm development in reducing VLA and ALMA data and the CASA software infrastructure. The array will have a progressive series of data products suitable to different users groups. The data products may also change based on how well supported a mode is - common modes will have higher level data products that add value to the user, while clearly not all permutations can benefit from such a degree of automation. 
As with the VLA and ALMA, the fundamental data product that will be archived are uncalibrated visibilities, enabling future reprocessing. The online software system will also produce flags to be applied to the visibilities that would identify known system problems such as antennas being late on source, or the presence of RFI. 

Automated post-processing pipelines will calibrate the raw data and create higher-level data products (typically image cubes) that will be delivered to users via the central archive. Calibration tables that compensate for large-scale instrumental and atmospheric effects in phase, gain, bandpass shapes, polarization and flux scale will be provided. Data analysis tools will allow users to analyze the data directly from the archive, reducing the need for data transmission and reprocessing at the user's institution. 

The VLA and ALMA ``Science Ready Data Products'' project will be an ngVLA pathfinder to identify common high-level data products that will be delivered to the Principal Investigator and to the data archive to facilitate data reuse. This model will also enable the facility to support a broader user base, possibly catering to astronomers who are not intimately aware of the nuances of radio interferometry, thereby facilitating multi-wavelength science.

\section{Overall System Performance}

The predicted performance of the array is summarized in Table \ref{tab:kpm}. This is an update to the performance estimates originally documented in \citet{2}\footnote{\url{http://ngvla.nrao.edu/page/refdesign}}.

The continuum and line rms values in Table \ref{tab:kpm} are for point source sensitivity with a naturally weighted beam. Imaging sensitivity is estimated based on a similar procedure as shown in \citet{3} and provided as a function of angular resolution in Table \ref{tab:imgsen}. The table is by necessity a simplification and the imaging sensitivity will vary from these reported values depending on the quality of the (sculpted) synthesized beam required to support the science use case. Herein, quality is defined as the ratio of the power in the main lobe of the sculpted beam attenuation pattern to the power in the entire beam attenuation pattern as a function of the FWHM of the synthesized beam \citep{4}. 

The brightness sensitivity of an array is critically dependent on the array configuration. The ngVLA has the competing aims of both good point source sensitivity at full resolution and good surface brightness sensitivity on a range of larger scales. Different array configurations that might provide a reasonable compromise through judicious weighting of the visibilities for a given application have been explored \citep{5} -- see \citet{6} for similar studies for the SKA. It is important to recognize that for any given observation, from full resolution imaging of small fields, to imaging structure on scales approaching that of the primary beam, some compromise will have to be accepted to enable a practical and flexible general purpose facility.

\begin{figure}[H]
\centering
\includegraphics[width=0.85\textwidth]{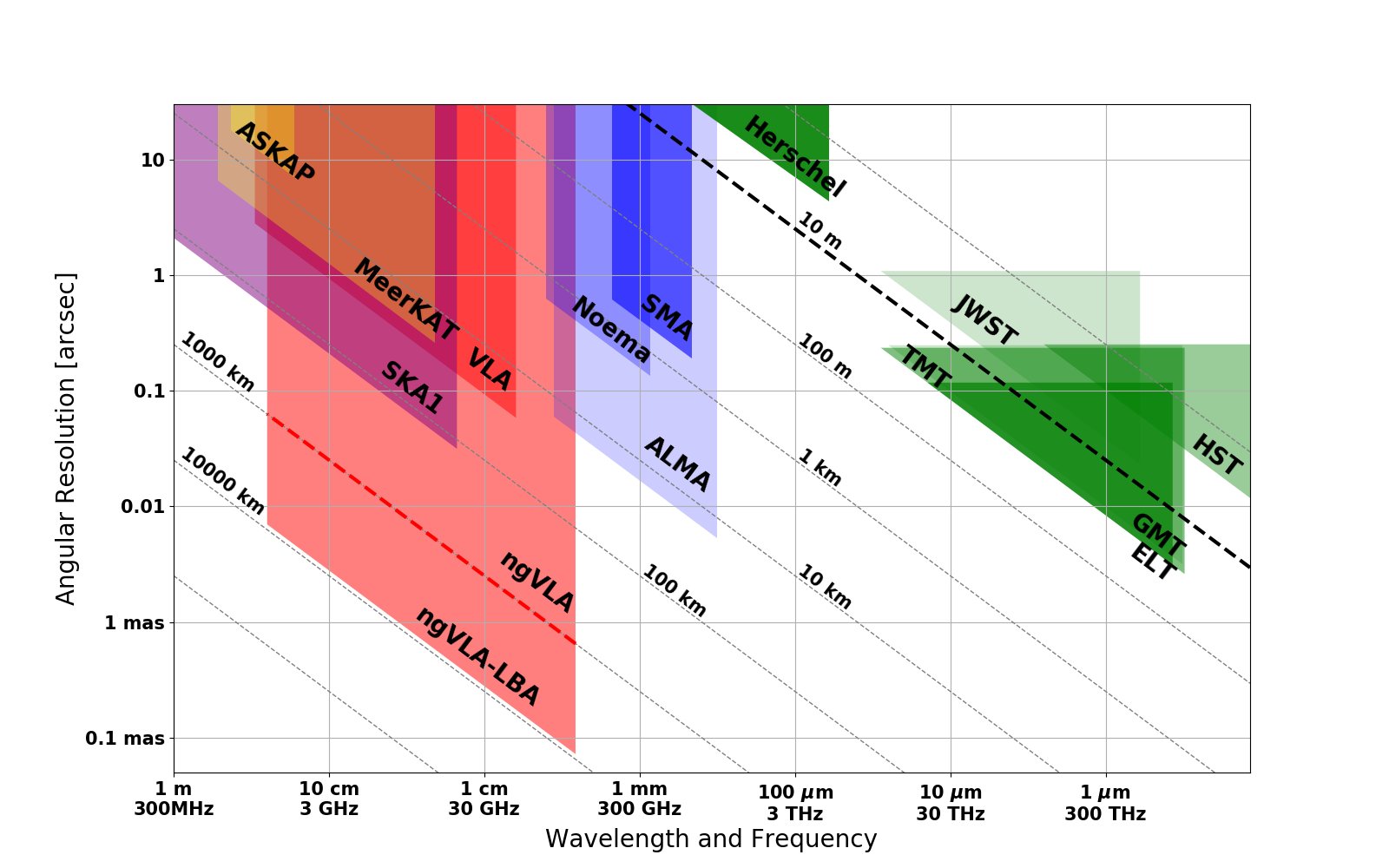}
\caption{\label{fig:SpatRes}Spatial resolution versus frequency set by the maximum baselines of the ngVLA as compared to that of other existing and planned facilities.}
\end{figure}

\begin{figure}[H]
\centering
\includegraphics[width=0.85\textwidth]{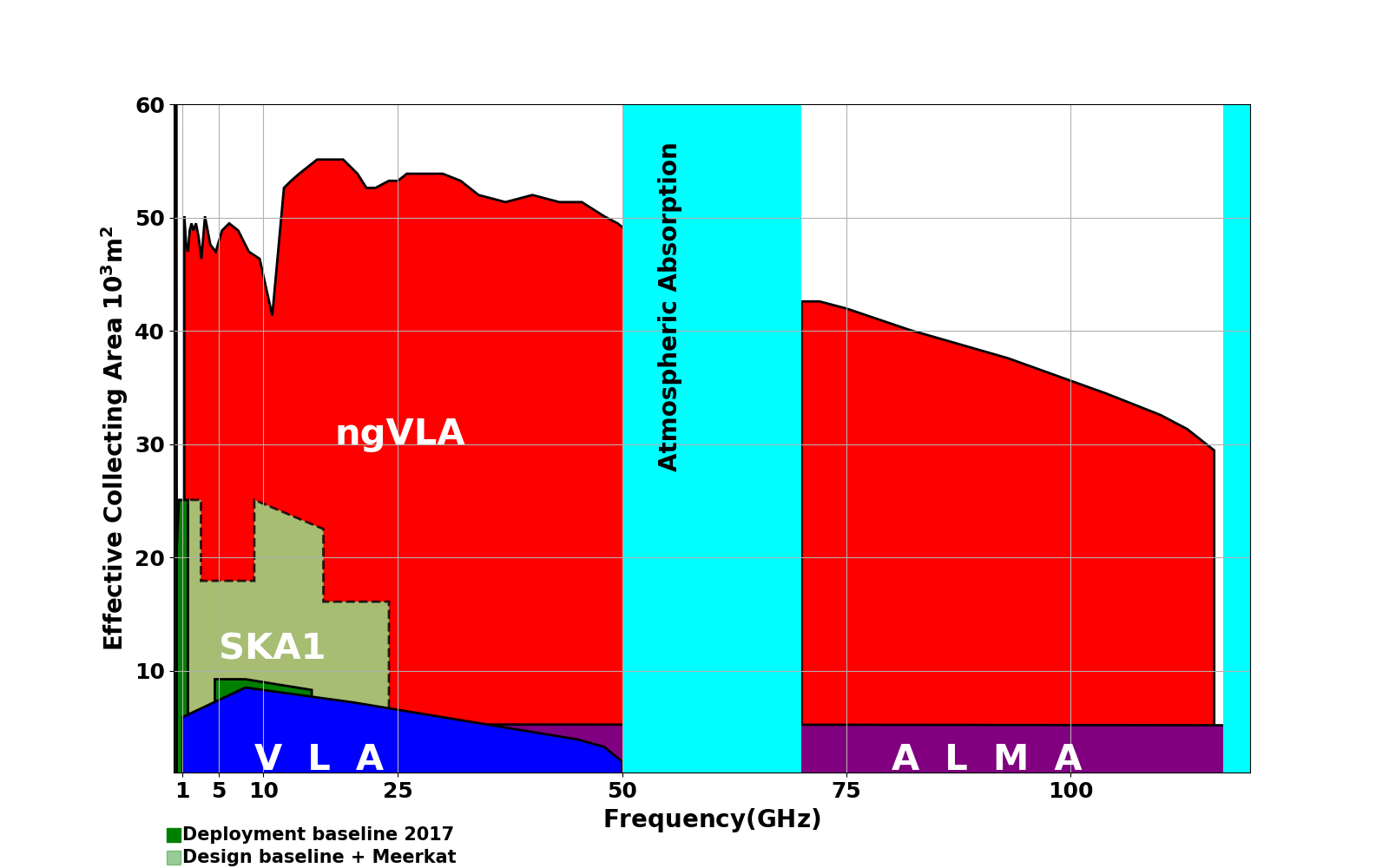}
\caption{\label{fig:Sensitivity}Effective collecting area versus frequency for the ngVLA as compared to that for other existing or planned facilities. Both the SKA1 `deployment baseline' (dark green) and `design baseline' (light green) are shown, inclusive of the MeerKAT array \citep{23}.}
\end{figure} 

Figure \ref{fig:SpatRes} shows a slice through the parameter space, resolution versus frequency, covered by the ngVLA along with other existing and planned facilities that are expected in the 2030s at radio to optical wavelengths. The maximum baselines of the ngVLA support a resolution of better than 0.5\,mas at 1\,cm. Coupled with the high sensitivity of the array, this resolution provides a unique window into the formation of planets in disks on scales of our own Solar system at the distance of the nearest active star forming regions.

Figure \ref{fig:Sensitivity} shows a second slice through parameter space: effective collecting area versus frequency. A linear-linear plot highlights the parameter space opened by the ngVLA. Note that the SKA-1 will extend to below 100\,MHz while ALMA extends up to almost a THz.
We note that there are other aspects of telescope phase space that are relevant, including field of view, mapping speed, surface brightness sensitivity, bandwidth, system temperature, dynamic range, etc. We have presented the two principle and simplest design goals, namely, maximum spatial resolution and total effective collecting area (as a gross measure of system sensitivity).

\begin{figure}[H]
\centering
\includegraphics[width=0.75\textwidth]{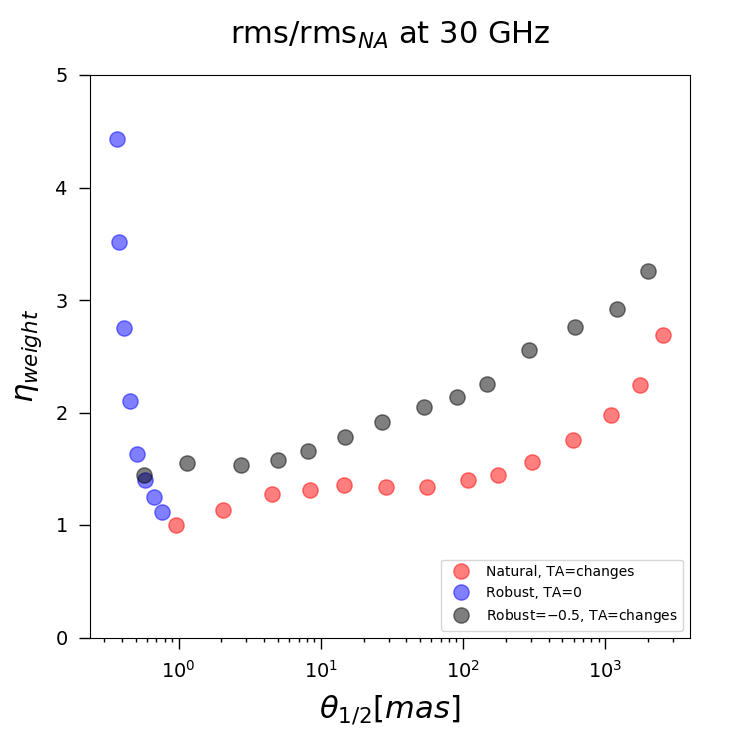}
\caption{\label{fig:inefficiency}   Image noise (rms)  at different angular resolutions (FWHM)  achieved by varying the imaging weights, simulated at 30 GHz.  The noise has been scaled relative to that of the naturally weighted image (rms$_{NA}$). The red symbols correspond to use of a ($u,v$)-taper and natural weights, and the blue symbols to Briggs robust weighting without a taper. The gray symbols are for Briggs robust = -0.5 and a varying ($u,v$)-taper, which has a large effect on beam quality (see Figure \ref{fig:psf_cuts}).}
\end{figure}

\begin{figure}[H]
\centering
\begin{tabular}{ccc}
\hspace*{\fill}%
\includegraphics[width=0.33\textwidth,  trim = 10 18 25 15, clip, angle = 0]{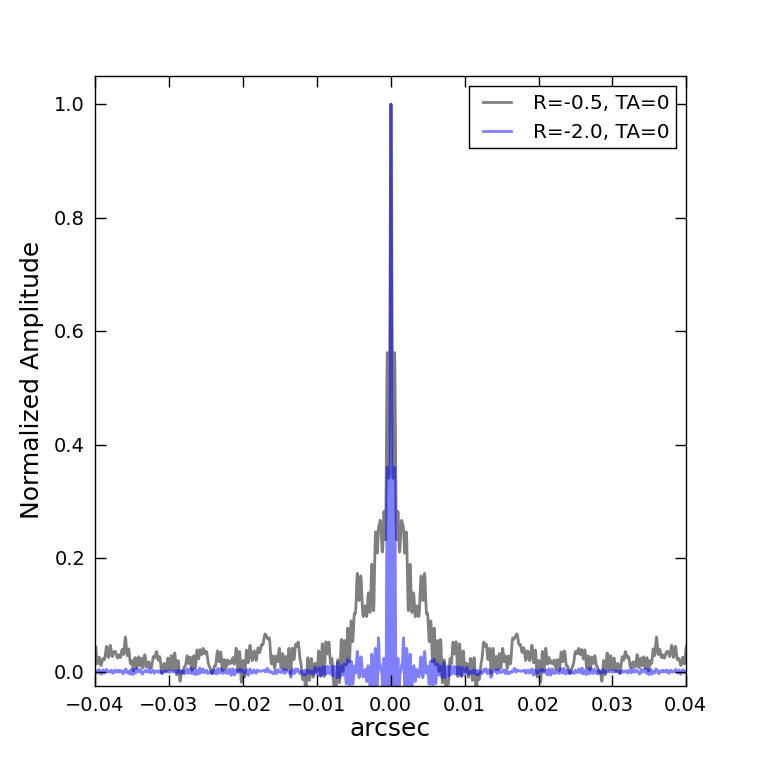}\hspace{-0.5cm}
&
\includegraphics[width=0.33\textwidth,  trim = 10 18 25 15, clip, angle = 0]{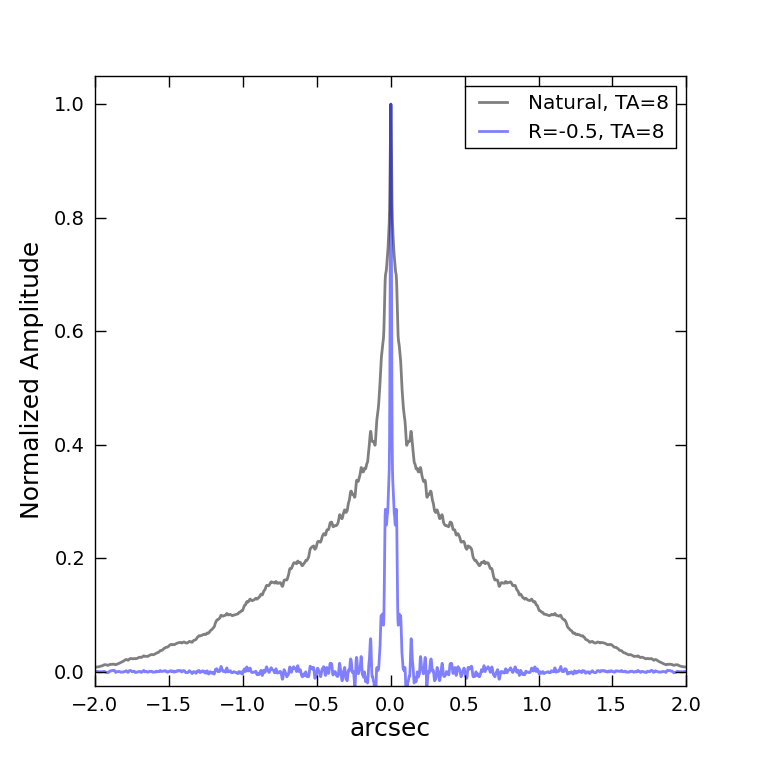}\hspace{-0.5cm} &
\includegraphics[width=0.33\textwidth,  trim = 10 18 25 15, clip, angle = 0]{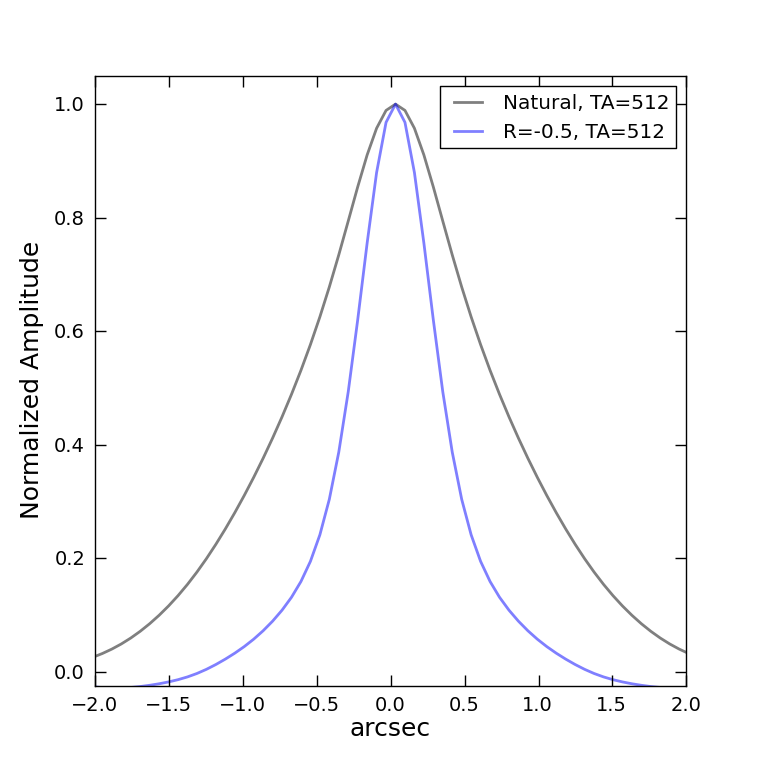}\\
\end{tabular}
\caption{
    Simulated 30 GHz PSFs over a range of resolutions, showing the effect of different imaging weights (TA: ($u,v$)-taper in mas, R:  Briggs robust parameter). The PSFs are a selection of the data presented in Figure~\ref{fig:inefficiency}: left panel (blue circles),  central and right panels (gray and red circles).  These examples illustrate how combinations of robustness and tapering allow for a beam of much higher quality at the expense of sensitivity.  }
\label{fig:psf_cuts}
\end{figure}

Imaging sensitivity will be dependent on the required resolution and imaging fidelity. Figures \ref{fig:inefficiency} and \ref{fig:psf_cuts} show the effects of adjusting imaging weights to vary the resolution and PSF quality. These figures are based on a 4 hour simulation at 30 GHz using the 244 antenna array configuration, for a source at +24$^{\circ}$ Declination observed during transit. The reported beam size is the geometric mean of the major and minor axes full width at half maximum (FWHM) of the synthesized beam as parameterized by Gaussian fitting in the CASA {\tt tclean} task. 
The highly centrally condensed antenna distribution leads to a naturally weighted beam that is not well characterized by a Gaussian function. Specific science applications may need to adjust the ($u,v$)-weighting and image parameters to `sculpt' a synthesized beam that is adequate for the particular science goal being considered \citep{3}. The results in Figures \ref{fig:inefficiency} and \ref{fig:psf_cuts} should be considered representative of the possibilities, and optimizing sensitivity vs. resolution will be a major area of investigation during telescope development.

In order to account for the change in sensitivity  due to use of imaging weights (relative to the naturally weighted rms $\sigma_{NA}$), we have adopted an efficiency factor  $\eta_{weight}$ such that the expected image rms after weighting is $\eta_{weight}\,\sigma_{NA}$.
The sensitivity calculations in Table~\ref{tab:imgsen}  include  $\eta_{weight}$,
estimated using the blue and red data series in Figure~\ref{fig:inefficiency} and by scaling $\theta_{1/2}$ with frequency ($\theta_{1/2} \times \nu / 30 \, \mathrm{GHz}$).

\begin{table}[H]
\centering
\begin{tabular}{L{3cm}|c|c|c|c|c|c|l}
\hline
Center Frequency [GHz] & 2.4 & 8 & 16 & 27 & 41 & 93 & Notes \\
\hline
\hline
Band Lower Frequency [GHz]	& 1.2	& 3.5	& 12.3	& 20.5	& 30.5	& 70.0	& a \\
Band Upper Frequency [GHz]	& 3.5	& 12.3	& 20.5	& 34.0	& 50.5	& 116.0	& a \\
Field of View FWHM [arcmin]	& 24.3	& 7.3	& 3.6	& 2.2	& 1.4	& 0.6	& b \\
Aperture Efficiency			& 0.77	& 0.76	& 0.87	& 0.85	& 0.81	& 0.58	& b \\
Effective Area, $A_{eff}$, x $10^3$ [m$^2$] & 47.8	& 47.1	& 53.8	& 52.6	& 50.4	& 36.0	& b \\
System Temp, $T_{sys}$ [K]		& 25	& 27	& 28	& 35	& 56	& 103	& a, e \\
Max Inst. Bandwidth [GHz]	& 2.3	& 8.8	& 8.2	& 13.5	& 20.0	& 20.0	& a \\
Sampler Resolution [Bits]	& 8		& 8		& 8		& 8		& 8		& 4		& \\ 	 
Antenna SEFD [Jy]			& 372.3	& 419.1	& 372.1	& 485.1	& 809.0	& 2080.5 & a, b \\
Resolution of Max. Baseline [mas]	& 2.91	& 0.87	& 0.44	& 0.26	& 0.17    & 0.07	& c \\
\hline
Continuum rms, 1~hr [$\mu$Jy/beam]	& 0.38	& 0.22	& 0.20	& 0.21	& 0.28	& 0.73	& d \\
Line Width, 10\,km/s [kHz]	& 80.1	& 266.9	& 533.7	& 900.6	& 1367.6	& 3102.1 \\	 
Line rms, 1 hr, 10 km/s [$\mu$Jy/beam]	& 65.0	& 40.1	& 25.2	& 25.2	& 34.2	& 58.3	& d \\ \hline
\end{tabular}
\caption{\label{tab:kpm}ngVLA Key Performance Metrics.
Notes:
(a)  6-band `baseline' receiver configuration.
(b) Reference design concept of 244 18\,m aperture antennas. Unblocked aperture with 160\,$\mu$m surface.
(c) Current reference design configuration, including LBA. Resolution in E-W axis.
(d) Point source sensitivity using natural imaging weights, dual polarization and all baselines (main array + LBA).
(e) At the nominal mid-band frequency shown. Assumes 1\,mm PWV at 93 GHz, 6\,mm PWV for other bands, 45$\deg$ elevation on sky.}
\end{table}

\begin{table}[H]
\centering
\begin{tabular}{L{4.5cm}|c|c|c|c|c|c}
\hline
\textbf{Center Frequency [GHz]}	& 2.4 & 8 & 16 & 27 & 41 & 93 \\
\hline
\hline
\textbf{Resolution [mas] : 1000} & & & & & & \\					
Continuum rms, 1 hr, Robust [$\mu$Jy/beam]	& 0.52	&0.34	&0.35	&0.39	&0.59	&2.24 \\
Line rms 1 hr, 10 km/s Robust [$\mu$Jy/beam]	& 88.9	&61.1	&43.3	&47.9	&70.9	&179.6 \\
Brightness Temp. ($T_{B}$) rms continuum, 1 hr, Robust [K] & 0.110 &6.4E-3	&1.7E-3	&0.7E-3	&0.4E-3	&0.3E-3\\
$T_{B}$ rms line, 1 hr, 10 km/s, Robust [K]	& 18.76	&1.16	&0.21	&0.08	&0.05	&0.03 \\ \hline
\textbf{Resolution [mas] : 100} & & & & & & \\				
Continuum rms, 1 hr, Robust [$\mu$Jy/beam]	& 0.50	&0.30	&0.27	&0.28	&0.40	&1.14 \\
Line rms 1 hr, 10 km/s Robust [$\mu$Jy/beam]	& 85.0	&53.6	&33.6	&34.8	&48.4	&91.3 \\
Brightness Temp. ($T_{B}$) rms continuum, 1 hr, Robust [K]	& 10.58	&0.56	&0.13	&0.05&0.03&0.02 \\
$T_{B}$ rms line, 1 hr, 10 km/s, Robust [K]	& 1794.1	&101.9	&15.9	&5.8	&3.5	&1.3 \\ \hline
\textbf{Resolution [mas] : 10} & & & & & & \\					
Continuum rms, 1 hr, Robust [$\mu$Jy/beam]	& 0.41 &	0.27	&0.26	&0.27	&0.38	&0.97 \\
Line rms 1 hr, 10 km/s Robust [$\mu$Jy/beam]	& 69.9 &	48.3&	32.4&	33.2&	46.3&	77.7 \\
Brightness Temp. ($T_{B}$) rms continuum, 1 hr, Robust [K]	& 870.58	& 50.51&	12.42&	4.53&	2.77&	1.36 \\
$T_{B}$ rms line, 1 hr, 10 km/s, Robust [K]	& 1.5E5	& 9173	&1540	&555&	335	&109 \\ \hline
\textbf{Resolution [mas] : 1} & & & & & & \\					
Continuum rms, 1 hr, Robust [$\mu$Jy/beam]	& - &	20.87	&0.31	&0.21	&0.29	&0.90 \\
Line rms 1 hr, 10 km/s Robust [$\mu$Jy/beam]	& - &	3789.8	&38.2	&25.7	&34.7	&72.0 \\
Brightness Temp. ($T_{B}$) rms continuum, 1 hr, Robust [K]	& -	& 4.0E5	&1466	&350&	207	&126 \\
$T_{B}$ rms line, 1 hr, 10 km/s, Robust [K]	& -	& 7.2E7	&1.8E5	&4.3E4	&2.5E4	&1.0E4 \\ \hline
\textbf{Resolution [mas] : 0.1} & & & & & & \\					
Continuum rms, 1 hr, Robust [$\mu$Jy/beam]	& - &	-	&-	&-	&-	& 20.96 \\
Line rms 1 hr, 10 km/s Robust [$\mu$Jy/beam]	& - &	-	&-	&-	&-	& 1683.2 \\
Brightness Temp. ($T_{B}$) rms continuum, 1 hr, Robust [K]	& -	& -	&-	&-&	-	& 2.9E5 \\
$T_{B}$ rms line, 1 hr, 10 km/s, Robust [K]	& -	& -	&-	&-	&-	& 2.0E7 \\ \hline
\end{tabular}
\caption{\label{tab:imgsen}Projected image sensitivity as a function of angular resolution. These calculations include $\eta_{weight}$ and are scaled by frequency, as described in the text.}
\end{table}


\acknowledgements The ngVLA is a project of the National Science Foundation's National Radio Astronomy Observatory which is operated under cooperative agreement by Associated Universities, Inc.  



\end{document}